\documentclass[12pt]{article}
\usepackage{graphicx}
\usepackage{amssymb,url, amsthm, amsmath, verbatim}
\newtheorem{theorem}{\noindent Theorem}

\author{A.~Vershik}

\begin{document}
\begin{center}{}\bf\textbf{THE BEHAVIOR OF LAPLACE TRANSFORM OF THE INVARIANT MEASURE
ON THE HYPERSPHERE OF HIGH DIMENSION.}
\bigskip
\bigskip

{}\bf \textbf{A.~M.~VERSHIK}
\end{center}
\rightline{\it To my friend Dima Arnold}
\date{28.03.08}


\begin{abstract}
We consider the sequence of the hyperspheres $M_{n,r}$ i.e. the
homogeneous transitive spaces - of the Cartan subgroup $SDiag(n,\Bbb
R)$ of the group $SL(n,\Bbb R), n=1 \dots$, and studied the
normalized limit of the corresponding sequence of the invariant
measures $m_n$ on those spaces. In the case of compact groups and
homogeneous spaces, as example - for classical pairs $(SO(n),
S^{n-1}), n=1 \dots$ - the limit of corresponding measures is the
classical infinite dimensional gaussian measure - this is well-known
Maxwell-Poincare lemma. Simultaneously that Gaussian measure is a
unique (up to scalar) invariant measure with respect to the action
of infinite orthogonal group $O(\infty)$. This coincidences means
the asymptotic equivalence between grand and small canonical
ensembles for the series of the pairs $(SO(n), S^{n-1})$. Our main
result shows that situation for noncompact groups, for example for
the case $(SDiag(n,\Bbb R),M_{n,r})$ (the definitions see below), is
completely different: the limit of measures $m_n $ does not exist in
literal sense, and we show that only normalized logarithmic limit of
the Laplace transform of those measures does exist. In the same time
there exists the measure which is invariant measure with respect to
continuous analogue of Cartan subgroup of the group $GL(\infty)$ -
this is so called infinite dimensional Lebesgue (see \cite{V}). This
difference is an evidence of the non-equivalence between the grand
and small canonical ensembles for the noncompact case.
\end{abstract}

\section{Setting of the problem.}

Consider the series of classical Lie groups $G_n$ and corresponding
homogeneous space $M_n$, equipped with $G_n$-invariant measures
$m_n$. Does there exist the natural (weak) limit of such sequences
of measures as a measure defined in some infinite-dimensional space?

There are at least two specifications of the question: we can try to
find a measure which is invariant under the action of an infinite
dimensional analog of our groups; and the second approach is to find
a direct limit of the finite dimensional measures. These two
approaches coincide for the classical case of orthogonal group
$SO(n)$ and spheres $S^{n-1}$ as homogeneous spaces- we obtain a
standard Gaussian measure in both cases - in the first approach - as
unique (up to the scalar) ergodic measure which is invariant under
infinite dimensional orthogonal group $SO(\infty)$ (Schoenberg
theorem see \cite{V,Ak}) and in the second approach -as
Maxwell-Poincare lemma (MP-lemma) about the weak limit of the
sequence of normalized Lebesgue measures on the $n-1$-dimensional
spheres of the radius $c\sqrt n$ (see detail discussion of this case
in \cite{V}).

 We consider the generalization of this scheme for noncompact group.
 The first nontrivial example is given by the case of the diagonal
 (Cartan) subgroup $SDiag(n,\Bbb R)$ of the group $SL(n \Bbb R)$. More
exactly, in the paper \cite{V} we considered the group of positive
diagonal matrices with determinant $1$,-  the group $SDiag_+(n,\Bbb
R)$, and its homogeneous space - the manifold, which sometimes
called "hypersphere of radius $r$":
$$M_{n,r}= \emph{} \{\{y_k\}:\prod_{k=1}^n y_k =r^n, \quad  y_k>0, k=1\dots n;
\}$$ as homogeneous transitive space. ($r=r_n$ depends on $n$). One
of the infinite-dimensional analogue of the group $SL(n \Bbb R)$ is
the group of all bounded linear operators on the space of Schwartz
distributions on a manifold $T$. Then the continuous analog of
diagonal (Cartan) group can be define as a group of multiplicators:
$M_a \xi(.)= a(.)\xi(.)$, where $\xi$ is a Schwartz distribution on
the manifold $T$ and $f$ run over the tame functions on $T$. Define
the multiplicative abelian group $\cal M$ of the {\it positive
functions} $a(.)$ with the condition:
$$\int_{t \in T} \ln a(t)dt =0,$$
this condition is the logarithmic analog of the condition $det A=1$
on diagonal matrices: \footnote{It is useful to consider more wide
group: ${\cal M}=\{a:\int_T \ln a(t)dt < \infty\}$ -this analogue of
the group $GDiag_+(n, \Bbb R)$}. Thus $\cal M$ is the continuous
analogue of the $SDiag_+(n,\Bbb R)$. It was proved (\cite{TVY}) that
there exist the one-parameter family ${\cal L}_{\theta} \quad
\theta>0 $ of $\cal M$-invariant measures on the space $S(T)$ of the
Schwartz distribution on the manifold $T$; here $S(T)$ is Schwartz
space; this family of the measures is unique up to scalar; the
measures (${\cal L}_1$ is infinite dimensional Lebesgue measure).

The Laplace transform of the measures ${\cal L}_{\theta}; \quad
\theta>0 $ is the functional
$$\Psi_{\theta}(f)\equiv \int_{\xi \in S(T)} e^{-<f,\xi>}
d{\cal L}_{\theta}(\xi)=\exp \{-\theta\int \ln f(t) dt\},$$ the
function $f$ is the positive tame function of the manifold $T$
(see\cite{V}). Note that we can regard the above formulas for the
Laplace transform as a definition of the measures ${\cal
L}_{\theta}$; a constructive definition and the detailed list of the
properties of the infinite dimensional Lebesgue measure ${\cal L}_1$
can be found in \cite{V}. Note, that the measure ${\cal L}_1$ (as
well as ${\cal L}_{\theta}$) can be considered as a low of "infinite
divisible process with sigma-finite measure" generated by semigroup
of sigma-finite measure on the $[0,\infty)$ with the density
$$\frac{x^{\theta-1}}{\Gamma(\theta)}, \theta>0.$$
This explains the formula above for Laplace transform of measure
${\cal L}_1$.

 In this paper we will calculate directly the asymptotics of the
Laplace transform of the invariant measures on the manifolds
$M_{n,r}$ and see that it gives another answer than in the first
method: the limit of Laplace transform of the measures $m_n$ does
not coincide with Laplace transform of measures ${\cal L}_{\theta}$.
Thus we have, in a sense,  the non-equivalence of two ensembles:
Lebesgue measure ${\cal L}_1$ is a natural measure on the grand
canonical ensemble with Laplace transform written above; from other
side the limit of measure on the small ensembles does not exist
literally. Nevertheless it is interesting to understand the behavior
of asymptotics of the Laplace transform of invariant measure on the
hypersphere - the function $L$, - which has very intriguing
properties. We briefly discuss this subject in the end of this
paper.

\section{The Laplace transform of measures on hyperspheres $M_{n,r}$}

We want to find the asymptotic properties of the invariant measure
on the positive Cartan subgroup $SDiag_+ (n, \Bbb R)$ (positive
diagonal real matrices) of the group $SL(n,\Bbb R)$ when $n$ tends
to infinity. More exactly we want to find the {\it Laplace transform
$D_n(.)$ of the invariant $\sigma$-finite measure on hypersphere}
$M_{n,r}$:

$$M_{n,r}=
\emph{} \{\{y_k\}: y_k>0, k=1\dots n; \prod_{k=1}^n y_k =r^n\},$$

(in general, the radius $r$ of the hypersphere  depends on $n$). We
use Laplace transform instead of Fourier transform in the case of
MP-lemma because the measure $m_n$ is not finite but sigma-finite.
The group $SDiag_+$ acts transitively and freely on this
hypersphere. So we want to investigate the following integral over
invariant measure $m_n(.)$ on $M_{n,r}$:

$$D_n(f)=\int _{M_{n,r}}
\exp\{-\sum_{k=1}^n f_k \cdot y_k\}dm_n(y).$$

By definition the functional $D_n(f)$ as function of vector
$f=(f_1,\dots f_n)$ {\it is Laplace transform} of the measure $m_n$
on the hypersphere ${M_{n,r}}$ . The measure $m_n(.)$ is the Haar
measure on the group $SDiag_+$. Since Haar measure is defined up to
constant we can choose the normalization later, but here the choice
of the coordinates $y$ has fixed by the normalization of Haar
measure. Our goal is to find the asymptotic property of measure
$m_n(.)$, and for this we use Laplace of the $\sigma$-finite
measure.

In this section we will calculate the asymptotic of that integral.
Let us change variables: $y_k\mapsto \frac{\rho_n(f)y_k}{f_k}$, ãäå
$\rho_n(f)=(\prod_{k=1}^n f_k)^{\frac{1}{n}}$ and then we obtain:
$$D_n(f)=\int_{M_{n,r}}
\exp\{-\rho_n(f) \sum_{k=1}^n y_k\} dm_n(y).$$

Let $y_k=e^{x_k},k=1,\dots n$, then we have
$$D_n(f)=\int_{P_n} \exp\{-\rho_n(f)\sum_{k=1}^n
\exp x_k\}\prod_k dx_k,$$ where  $P_n=\{(x_1 \dots x_n):\sum_k
x_k=n\ln r_n\}$.

 Finally letting $x_k\mapsto x_k -\ln r_n$,
 we obtain the following expression for:
$$D_n(f)=$$
$$=\int_{{\Bbb R}^n} \exp{\{-\rho_n(f) r_n\sum_{k=1}^n
   e^{x_k}\}}\delta_0(\sum_{k=1}^n x_k)\prod_{k=1}^n dx_k\equiv$$
   $$=\int_{H_n}\exp\{-\rho_n(f) r_n \sum_{k=1}^n
   e^{x_k}\}dx, $$
 where integration runs over hyperplane:
$$H_n=\{(x_1,\dots x_n):\sum_k x_k=0\},$$

and
 $\rho_n(f)=(\prod_{k=1}^n
f_k)^{\frac{1}{n}}$.

Note that keeping in the mind the continuous limit of whole picture
we can regard vector $f=(f_1, \dots f_n$ as the values of function
$f(.)$ at some points $t_k, \quad f_k=f(t_k); k=1 \dots n$ of the
domain of the functions $f$, then functional $\rho(f)$ becomes as
functional $\exp{\int \ln f(t)dm(t)}$, which is defined on the class
of positive functions $f$ of continuous argument $t$ with finite
integral of logarithm: $\int \ln f(t)dt,\infty$ (see details in
\cite{V}). In any case,  functional $D_n$ depends on geometrical
average $\rho(f)$ of the coordinates of the vector $f$ only.

Now recall the definition of the function $F_n$ on the positive half
line ${\Bbb R}_+$ which sometime is called Mellin-Barnes functions
(see f.e.\cite{Par}):

 $$F_n(\lambda)=\int_{H_n}\exp \{-\lambda \sum_{k=1}^n \exp x_k\}dx,$$ where
 $H_n=\{(x_1,\dots x_n)\in {\Bbb R}^n:\sum_{k=1}^n x_k=0\}$.

 We have:
 $$D_n(f)=F(\rho_n(f)r_n),$$

 and this formula {\it reduces our problem to the calculation of the asymptotics of
 the function  $F_n(.)$} when $n$ tends to infinity. The role of multiplier $\rho$
 will be discussed later.
 In some sense we get an asymptotic problem about function of one
variable ('radius") instead of many variables -the explanation lies
in the high symmetry of the initial problem under the group
$SDiag(n,{\Bbb R}_+)$. We have observed same effect in the problem
about asymptotics of the uniform measures on the spheres which also
reduces to the asymptotics of the Bessel functions $J_n(.)$ of one
variable (see \cite{Ak,V}). The function $F_n$ plays here the same
role and Bessel function in the compact case. But the answer as we
will see will be very different.

As is well-known, the function $F_n$ satisfies to the differential
equation:
$$(1+\lambda\frac{d}{d\lambda})^{n-1}\frac{dF_n}{d\lambda}=F_n(\lambda),$$
and related to the hypergeometric functions. We do not use this
properties of $F_n$.

\section{Calculations of the Asymptotics of the Function $F_n$}

 First of all find integral
 $$\int_0^{\infty}F_n(e^{-\frac{t}{n}})\cdot e^{-ts}dt=$$
 $$=\int (n) \int e^{-e^{-\frac{t}{n}}
 \cdot [e^{x_1}+e^{x_2}+\dots +e^{x_n}]} \cdot e^{-ts}dx_1 \dots
 dx_{n-1}dt=$$

 where the integration over $$H_n\times \Bbb R=\{(x_1,\dots x_n,t)\in {\Bbb
 R}^{n+1}: \sum_{k=1}^n x_k=0, t>0 \}.$$

 $$=\int (n)\int \exp [-\sum_{k=1}^n e^{({x_k-\frac{t}{n}})}] \cdot e^{-ts}dx_1\dots dx_{n-1}dt.$$

 Change variables: $(x_1,\dots x_{n-1},t) \mapsto (x_1\dots x_n)$,
 where
 $x_n=t-x_1-\dots -x_{n-1},\quad \mbox{ò.å.}$, so $t=\sum_{k=1}^n
 x_k$, and let $$y_k=\exp (x_k-\frac{t}{n}), \quad k=1 \dots n.$$

 The integral under consideration equals:
 $$=\int \cdots (n)\cdots \int e^{-\sum_{k=1}^n y_k} \prod_{k=1}^n (y_k)^s\frac{dy_k}{y_k}=$$
 $$= \prod_{k=1}^n [\int_0^{\infty}e^{-y_k}y_k^{s-1}]= \Gamma(s)^n$$

 This, $$\int_0^{\infty}F_n(e^{-\frac{t}{n}})\cdot
 e^{-ts}dt=\Gamma(s)^n.$$

 Now make the change the variable $e^{-\frac{t}{n}}=\lambda$; then we
 have:

  $$\int_0^{\infty}n F_n(\lambda)\lambda^{ns-1}d\lambda= \Gamma (s)^n.$$
  {\it So the function $F_n(.)$ is the inverse Mellin transform of $n$-th
 degree of Gamma-function- $\Gamma(s)^n$} (up to multiplier $n$ which we will use later).
 Thus the functions $F_n(\lambda)$ and
 $\Gamma(s)^n$ form so-called "Mellin's pair".

 For example, for n=1 the Mellin pair is the pair of functions $(\exp s,\Gamma(s))$.

 Now in order to express function  $F_n(.)$ and to find its
 asymptotics we need to use the standard calculation of the
 following contour integral and apply the saddle point method:

 $$F_n(\lambda)=\frac{1}{2\pi in}\int_{\gamma-i\infty}^{\gamma+i\infty} [\Gamma(s)]^n \lambda^{-ns}ds.$$

 This is the Laplace transforms of the sequence of measures on the orbit of
 $S_+Diag(n,\Bbb R)$. We want to study the asymptotics of this sequence when $n \to \infty$.

 Rewrite the last formula in the "real" form and put $s=\gamma + it$

 $$F_n(\lambda)=\frac{1}{2\pi in}\int_{-\infty}^{+\infty} [\Gamma(\gamma+it)]^n \lambda^{-n(\gamma+it)}dt.$$

Now we are ready to apply the saddle-point method to the calculation
of the asymptotics of the function $F_n$ and first of all to find
saddle point.

\section{Saddle point method for $F_n$.}

 It turns out that the suitable saddle point is the point $(\gamma,0)
 \in \Bbb C$ where $\gamma$ is the root of the following equation:

  $$\ln \lambda=\frac{\Gamma'(\gamma)}{\Gamma(\gamma)},$$
 or
  $$\lambda=\exp\bigg\{\frac{\Gamma'(\gamma)}{\Gamma(\gamma)}\bigg\}.$$

 Recall that the fraction $$\frac{\Gamma'(\gamma)}{\Gamma(\gamma)}$$ or
 logarithmic derivative of $\Gamma$-function is well-known
 Euler's $\psi$-function and its properties are well-known (see f.e.
 \cite{Er,RGr}).

  In particular $\psi$-function is
 strictly monotonic on the half-line $(0,\infty)$ and maps it to
 the whole line $(-\infty,+\infty)$, so
 for all $\lambda>0$ there exists  a value $\gamma=\gamma(\lambda)$ which
 satisfies to the equation above.
 Thus $\gamma=\gamma(\lambda)$, as a function of $\lambda$, is inverse function to
 $\psi$-function. We will consider this function later.

   Rewrite our integral in the form:

 $$F_n(\lambda)=\frac{1}{2\pi
 i}\frac{\Gamma(\gamma)^n}{\lambda^{n\gamma}}\int_{-\infty}^{+\infty}
 \bigg\{\frac{\Gamma(\gamma+it)}{\Gamma(\gamma)}\bigg\}^n \cdot
 \lambda^{int}dt.$$
 Now expand the fraction under the integral into a power series
 in $t$ at a neighborhood of the point $\gamma$:

 $$\frac{\Gamma(\gamma+it)}{\Gamma(\gamma)}\cdot
 {\lambda}^{-it}=(1+\frac{\Gamma'(\gamma)}{\Gamma(\gamma)}it-$$
 $$-\frac{\Gamma''(\gamma)}{\Gamma(\gamma)}\cdot t^2/2 +\dots)(1-it\ln \lambda -t^2/2
 (\ln \lambda)^2+\dots )=$$

 Our choice of parameter $\gamma$ leads to the vanishing of the
 coefficient of the first power of $t$, so that we have
 $$= 1-\frac{t^2}{2}\bigg \{\frac{\Gamma''(\gamma)}{\Gamma(\gamma)}-(\frac{\Gamma'(\gamma)}
 {\Gamma(\gamma)})^2\bigg\}+\dots.$$

 Denote
 $$\sigma=\frac{\Gamma''(\gamma)\Gamma(\gamma)-\Gamma'(\gamma)^2}{\Gamma(\gamma)^2}.$$

 Then

 $$F_n(\lambda)=\frac{1}{2\pi n}\bigg\{\frac{\Gamma(\gamma)}{\lambda^{\gamma}}\bigg\}^n
 \int_{-\infty}^{+\infty} \bigg\{1-\frac{\sigma
 t^2}{2}+o(t^2)\bigg\}^n dt
 $$
 and, applying the arguments usual for saddle point method (check
 that point $(\gamma,0)$ is indeed the saddle point with respect to
 imagine and real axis) we obtain the asymptotics
 $$F_n(\lambda)\approx \frac{1}{2\pi n}\bigg(\frac{\Gamma(\gamma)}{\lambda^{\gamma}}\bigg)^n
 \int_{-\infty}^{+\infty}\exp\{-\frac{\sigma n t^2}{2}\}dt=
 \frac{L(\lambda)^n}{\sqrt {2\pi n \sigma^{-1}}},$$ where
 $$L(\lambda)=\frac{\Gamma(\gamma)}{\lambda^{\gamma}},\quad
 \lambda=\exp\bigg\{\frac{\Gamma'(\gamma)}{\Gamma(\gamma)}\bigg\}.$$

\section{The function $L$ and its behavior at zero and at infinity}
 Let us begin with the theorem-definition:
 \begin{theorem}
 Define the function $L$:
 $$L(\lambda)\equiv \lim_n\frac{\ln F_n(\lambda)}{n}.$$

 Then
 $$L(\lambda)=\frac{\Gamma(\gamma(\lambda))}{\lambda^{\gamma}},$$ where $\gamma$
 and $\lambda$  are satisfy to the equation:
 $$\lambda=\exp\bigg\{\frac{\Gamma'(\gamma)}{\Gamma(\gamma)}\bigg\}.$$
 \end{theorem}

It seems that the function $L$ is very interesting object; it looks
like free energy in statistical mechanics; author do not know if it
has been studied in the literature. This function plays role
analogous to the role of generators in the theory of semigroups
under convolution of the measures, or more exactly, the role
analogous to Fourier transformation of generator.

 Now we can look at the behavior of the function $L$,
which is the main goal of our calculations.

 Accordingly to the theorem 1 the function $L$
 can be considered as a function not only of $\lambda$ but which depend on $\gamma$:

$$L(\lambda)= \frac{\Gamma(\gamma(\lambda))}{\lambda^{\gamma(\lambda)}},$$

Consequently, first of all we need to clarify the interrelation
between variables $\lambda$ and $\gamma$ by the above equation for
$\gamma$ which can be rewritten as
 $$\lambda=\exp{\psi(\gamma)}.$$

\begin{theorem}
The asymptotical correspondence  between variables $\lambda$ and
$\gamma$ at infinity is the following:

$$\lambda=\gamma +o(1)$$

and the asymptotical correspondence between $\lambda$ and $\gamma$
at the point $0$ is the following:

$$\lambda(\gamma)= e^{\textbf{C}-\frac{1}{\gamma}}(1+o(1)),$$

or inversely:

$$\gamma\equiv \gamma(\lambda)
=\frac{1}{|\ln \lambda|+\textbf{C}+o(\lambda)}, \quad \lambda \sim
0,$$

here $\textbf{C}=0.577...$ is the Euler constant.
\end{theorem}
See graph of the function $\lambda=\lambda(\gamma)$ below.

\includegraphics[width=10cm,height=14cm,angle=270]{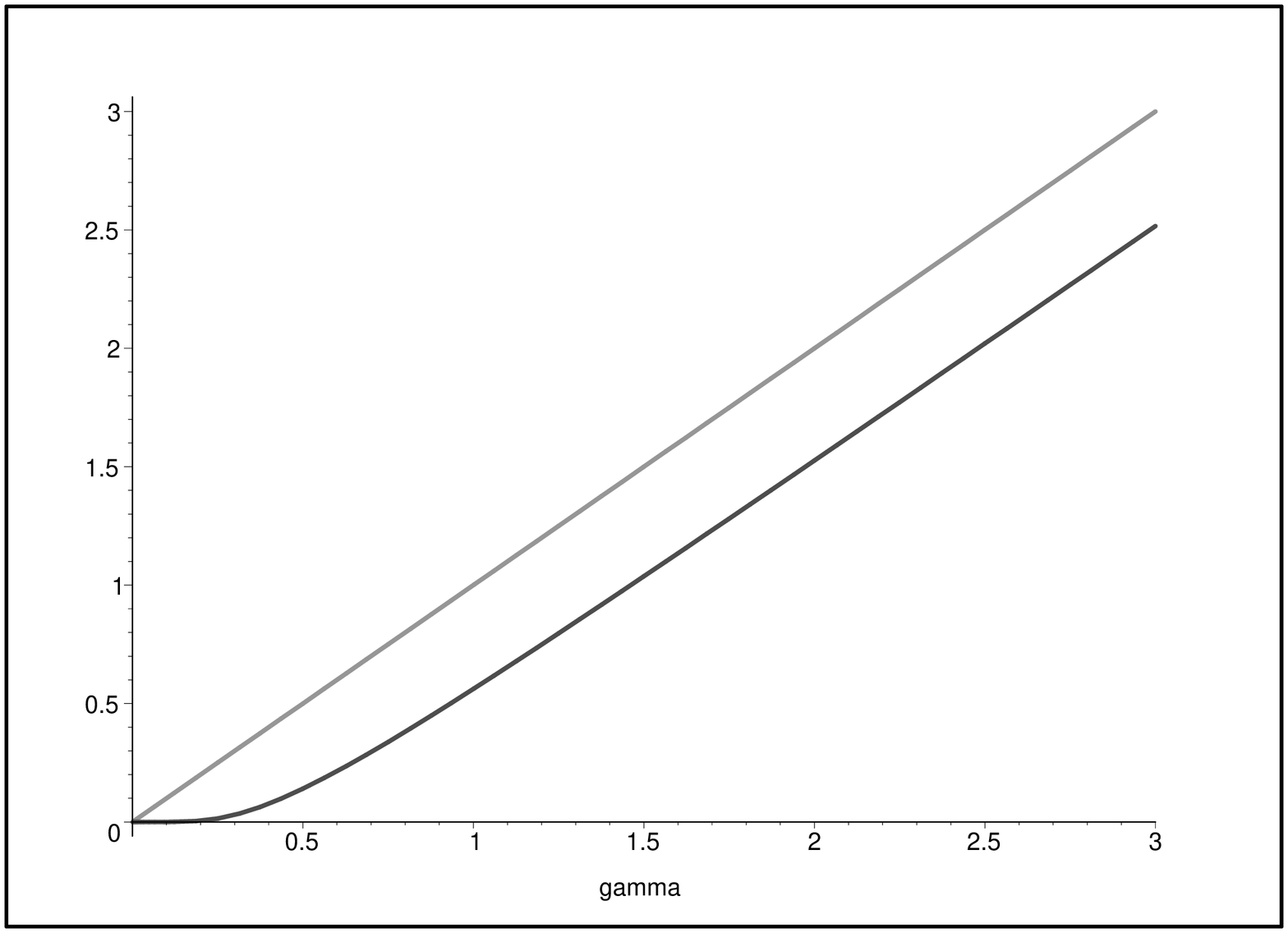}

\begin{proof}

We will use the relations for $\psi$-function (see \cite{RGr} 8.365)
$$\lim_{z\to \infty}(\psi(z+y)-\ln z)=0\quad \psi(z+1)=\psi(z)+\frac{1}{z}.$$
It follows that:
$$\psi(z)=\ln z +1/z+o(1/z),$$
as $z\to \infty$
 Since $\lambda =\exp{\psi(\gamma)},$ we have
$$\lambda =\exp \{\ln \gamma +o(1)\} =k_\gamma \cdot \gamma =\gamma
+o(\gamma^{-1});$$
 here $k_\gamma \to 1$ when $\gamma \to \infty$.

Now use the second relation at the neighborhood of $0$; then for $z
\sim 0$ we have
$$\psi(z)=\psi(1+z)-\frac{1}{z}=\textbf{C}+o(z)-\frac{1}{z},$$

Consequently $\lambda$ as function of $\gamma$  is the following:
$$\lambda \equiv \lambda(\gamma)=\exp\{\textbf{C}\}\cdot
\exp\{-\frac{1}{\gamma}\}(1+O(\gamma).$$

From this expression it is easy to obtain expression of $\gamma$ as
function of $\lambda$ which was done in the formulation of the
theorem.
\end{proof}

As we have seen, the asymptotical correspondences between $\lambda$
and $\gamma$ at zero and at infinity are very different:  at
infinity the variables are almost equal, while the correspondence at
a neighborhood of zero is very nontrivial: all jets of $\lambda$ as
function of $\gamma$ at zero are vanish.

We consider behavior  of function $L$ at zero and at infinity, and
also we are interested with value $1$. One can verify $L$ is
monotonically decreases on $(0,\infty)$ from $\infty$ at zero to $0$
at infinity.

As to asymptotic at $0$, our formulas and the properties of
functions $\Gamma(.)$ and $\psi(.)$ imply

\begin{theorem}
The behavior of the function $L$ at zero is the following:
$$\exp L(\lambda)=\frac{\textbf{C}}{\lambda}(1+o(\lambda)),$$
or
$$L(\lambda)=\ln \textbf{C}-\ln \lambda +o(\lambda),$$
and
$$\lim_{\lambda \to 0}\lambda \cdot e^{L(\lambda)} =\textbf{C} >0$$

The behavior of the function $L$ at infinity is the following:

$$L(\lambda)\cong \exp\{-\lambda\}, \quad \lambda \gg 0.$$
\end{theorem}

See graph of the function $L=L(\lambda)$ below.

\includegraphics[width=14cm,height=10cm,angle=0]{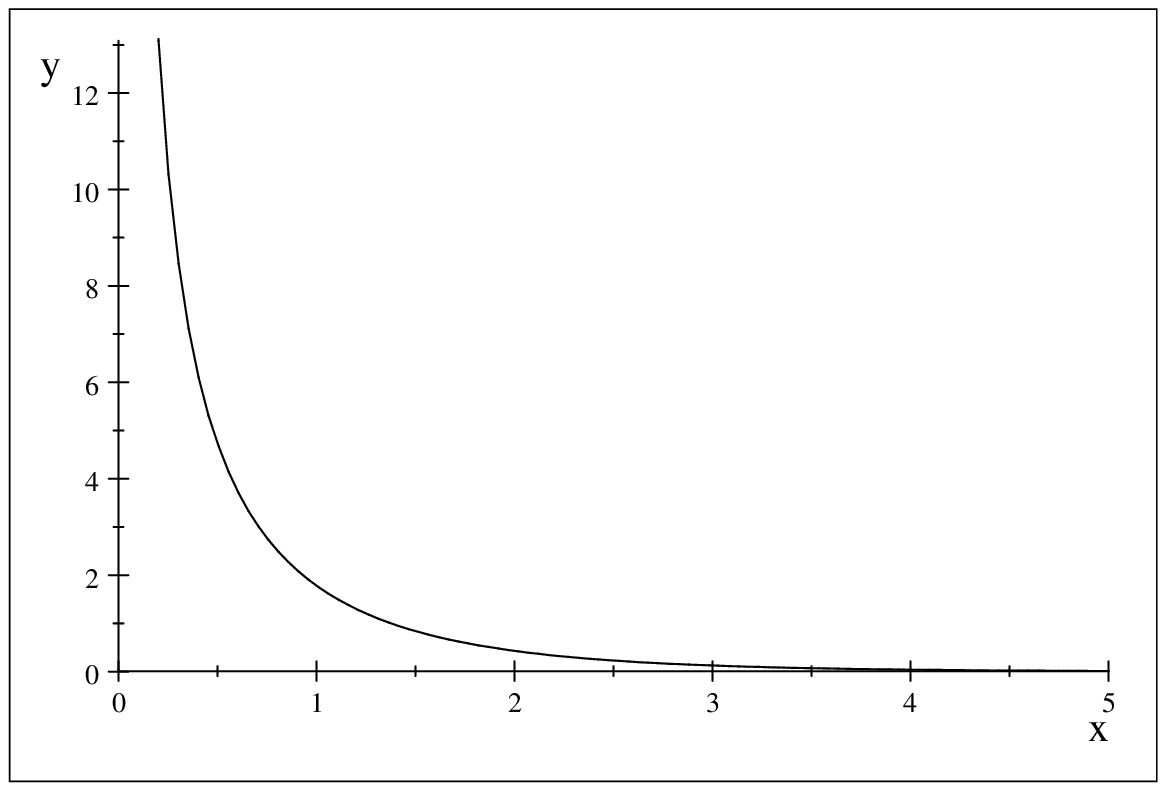}

\begin{proof} In order to find asymptotics of $L(.)$ at infinity
we apply to the formula of function $L$ the Stirling formula for
$\Gamma$-function,
$\Gamma(z)=z^{z-\frac{1}{2}}e^{-z}\sqrt(2\pi)(1+o(1)$, to the
formula for $L$, and use the relation $\lambda =\gamma +o(\lambda)$
at infinity. Then we have:

$$L(\lambda)=\frac{\Gamma(\gamma(\lambda))}{\lambda^{\gamma(\lambda)}}\approx
\frac{\Gamma(\lambda)}{\lambda^{\lambda}}\approx
\frac{\sqrt{2\pi}e^{-\lambda}}{\sqrt{\lambda}}$$

The asymptotical formula for the function $L$ at zero follows from
the fact that $\Gamma$-function has simple pole at zero with
residual equal to $1$, and Theorem 3:
$$L(\lambda)=-\ln\lambda+\textbf{C}+o(\lambda).$$
 In order to find the value $1$ of the function $L$ it is more
convenient to consider the function ${\bar
L}(\gamma)=L(\lambda(\gamma))$. Then

$${\bar L}(\gamma)=\frac{\Gamma(\gamma)}{\lambda(\gamma)^{\gamma}}.$$

Denote by $\gamma_{cr}$ the value of $\gamma$ for which $(\bar
L)(\gamma_{cr})=1$ (critical value). Using expression for
$\lambda=\exp \psi(\gamma)$ we see that if ${\bar L}=1$, (or, when
$\gamma=\gamma_{cr}$) then we have:

  $$\Gamma(\gamma_{cr})=\exp \gamma_{cr} \psi(\gamma_{cr}).$$

 In logarithmic scale we have the equation for this critical value:

$$\Gamma(\gamma_{cr})\ln \Gamma_{cr}(\gamma_{cr})-\gamma_{cr} \Gamma'(\gamma_{cr})=0.$$
It happened that there is a unique such $\gamma_{cr}$ and
approximately $$\gamma_{cr}\thicksim 1,38\dots;$$ the corresponding
value of (recall that
$\lambda=\exp\{\frac{\Gamma'(\gamma)}{\Gamma(\gamma)}\}$)
$$\lambda_{cr}\thicksim 0,95\dots.$$
\end{proof}
Now we can summarize the obtained information on the function $L$ in
order to find the asymptotics of the functions $F_n(.)$ in $n$. A
special role plays the behavior of $L$ in the neighborhood of
critical point.

\begin{theorem}
For each $\epsilon>0$ the values of function $F_n(\lambda)$
exponentially tends to infinity for all positive $\lambda <
\lambda_{cr}-\epsilon$, and exponentially tends to zero for $\lambda
> \lambda_{cr}+\epsilon$ when $n$ tends to infinity; there exist a sequence $\{\lambda_n\}$
which tends to $\lambda_cr$ such that
$$\lim_n F_n(\lambda_n)=1$$.
\end{theorem}

\section{Discussion}

We have calculated the asymptotic behavior of the Laplace transform
$D_n(f)$ of the invariant measures $m_n$ on the hyperspheres
$M_{n,r_n}$ which is defined by asymptotics of the function $F_n$.
The answer is given in terms of the distinguish function $L$ and
that analysis shows that there is no weak convergence of the
measures $m_n$ (in the sense of convergence of Laplace transform),
and consequently, it is impossible to say that it coincides with
Laplace transform of infinite-dimensional Lebesgue measure ${\cal
L}_1$.

Recall that the formula of Laplace transform of the measure $m_n$
$$D_n(f)=F(\rho_n(f)r_n),$$ involves the multiplier $\rho_n(f)$
which is the characteristic of the normalization of the hypersphere
$M_{n,r_n}$. The choice of this parameter has geometric sense - this
is the growth of radius $r_n$ of the hypersphere $M_{n,r_n}$.

We can conclude our analysis with the following words - {\it it is
impossible to choose the normalization of the radius of the
hyperspheres $r_n$ for which we obtain the non-degenerated limit
distribution of the invariant measures on the hyperspheres.} For
comparison in the compact case (Euclidean spheres) we can choose the
normalized radius equal to $c \sqrt n$ (MP-lemma).

We can say that there is no equivalence between grand- and micro-
canonical ensembles in the situation with the last one are
noncompact. We will discuss this phenomena elsewhere.

At the same time the analysis of the function $L$ itself is very
important itself, because it shows that the behavior of the Laplace
transforms $F_n$ in $n$ asymptotically looks like Laplace transform
of the powers (in the sense of convolution) of some distribution.
The main question is: whether function $L$ stands for a generator of
some semigroup of infinite-divisible distributions on the half-line,
or not.

In conclusion we can say that there is still no suitable theory of
divisibility for Laplace transform of the sigma-finite distributions
as well as Levy-Khinchin type formula for generators and so on, (at
least up to the authors knowledge of subject).

 Our result also shows although the groups $SDiag(n,\Bbb R)$
 in the natural sense tends (when $n$ tends to infinity)
 to the group $\cal M$ of all symmetries of that Lebesgue measure
 ${\cal L}_1$, - it can happened that the individual ergodic theorem
 (or law of large number) for the increasing sequences
 of the groups $\{SDiag(n,\Bbb R), n=1\dots \}$ not valid.

\textbf{Acknowledgment}. Professor M.Graev point out to some
properties of the functions $F_n(\lambda)$. Professor D.Zagier
recommended me to use saddle point method for calculation of the
function $L$. Dr.F.Petrov noticed that the function $L$ can be also
represented as the exponent of the Legeandre transform of the
logarithm of Gamma-functions $\ln \Gamma(.)$; this gives another
possibility for calculation of the function $L$ however the
justification of the procedure is more complicate in this case than
in the case of saddle point method. Dr.D.Finkelstein and Prof.
A.Borodin made some graphs of the functions $L$ and $\lambda$. Prof.
N.Tsilevich made several editorial remarks on the last version of
the paper. The author thanks all of them for the help. The author
also thank the Mathematical Scientific Research Institute at
Berkeley, where this article was finished during the semester
"Combinatorial Representation Theory" (Spring 2008),
RFBR-08-01-000379a, and the President grant NSh-2460.20081112.

\end{document}